\documentclass[aps,apl,twocolumn]{revtex4-1}
\usepackage{graphicx}
\usepackage{dcolumn}
\usepackage{bm}

\begin{document}

\preprint{}

\title{Spectrally-resolved hyperfine interactions between polaron and nuclear spins in organic light emitting diodes: Magneto-EL studies}

\author{S. A. Crooker$^1$, F. Liu$^2$, M. R. Kelley$^1$, N. J. D. Martinez$^1$, W. Nie$^1$, A. Mohite$^1$, I. H. Nayyar$^1$, S. Tretiak$^1$, D. L. Smith$^1$, P. P. Ruden$^2$}

\affiliation{$^1$Los Alamos National Laboratory, Los Alamos, NM 87545  USA}
\affiliation{$^2$University of Minnesota, Minneapolis, MN 55455  USA}

\date{\today}
\begin{abstract}
We use spectrally-resolved magneto-electroluminescence (EL) measurements to study the energy dependence of hyperfine interactions between polaron and nuclear spins in organic LEDs.  Using layered devices based on Bphen/MTDATA -- a well-known exciplex emitter -- we show that the increase in EL emission intensity $I$ due to small applied magnetic fields of order 100~mT is markedly larger at the high-energy blue end of the EL spectrum ($\Delta I/I \sim$11\%) than at the low-energy red end ($\sim$4\%). Concurrently, the \emph{widths} of the magneto-EL curves increase monotonically from blue to red, revealing an increasing hyperfine coupling between polarons and nuclei and directly providing insight into the energy-dependent spatial extent and localization of polarons.
\end{abstract}

 \maketitle
Several recent experiments have shown that small applied magnetic fields \textbf{B} on the order of 10-100~mT can induce substantial ($\sim$10\%) changes in the total light intensity emitted by organic light-emitting diodes (OLEDs) \cite{KalinowskiCPL2003, Bussman, Sheng, Iwasaki, Odaka, Wu, Desai, FJWang, BinHu, NguyenNatMat, KerstenPRL, Ehrenfreund, NguyenPRB}.  While initially surprising in view of the fact that the polymers and small molecules used in OLEDs are primarily composed of non-magnetic atoms (H, C, N), it was quickly appreciated that \emph{hyperfine} spin interactions underpinned these phenomena.  Specifically, the coupling of the electron and hole polaron spin to the many nuclear spins in the host material generates randomly-oriented local effective magnetic fields about which electron and hole polaron spins can precess.  This precession leads to spin mixing between singlet and triplet polaron-pair states, which are precursors to exciton or exciplex formation in an OLED.  Applied fields \textbf{B} suppress this hyperfine-induced mixing, altering the population balance between singlet and triplet excitons or exciplexes, which in turn modifies the electroluminescence (EL) efficiency.

The detailed dependence of EL intensity on \textbf{B} allows direct insight into not only the rates of singlet and triplet exciton/exciplex formation, but also reveals the strength of hyperfine coupling and therefore provides a measure of the spatial extent (size) of the electron and hole polarons.

In magneto-EL studies to date \cite{KalinowskiCPL2003, Bussman, Sheng, Iwasaki, Odaka, Wu, Desai, FJWang, BinHu, NguyenNatMat, KerstenPRL, Ehrenfreund, NguyenPRB}, only the total (spectrally-integrated) EL intensity was measured as a function of \textbf{B}. However, OLED emission spectra typically span a very broad wavelength range, reflecting the fact that excitons and exciplexes form over a wide range of energies, and with varying degrees of localization for which different hyperfine couplings may be expected.  Here we spectrally resolve the magneto-EL from MTDATA/Bphen OLEDs - a well known exciplex emitter - and show that the increase in EL intensity due to \textbf{B} is significantly larger at the high-energy blue side of the spectrum than at the low-energy red side.  Most importantly, the widths of the magneto-EL curves increase by over a factor of two from blue to red, directly revealing an increasingly strong hyperfine coupling and providing insight into the energy-dependent spatial extent and localization of the emitting states.

Figure 1(a) depicts our layered OLED structures. They have transparent ITO/PEDOT anode contacts (indium tin oxide/ polyethylenedioxythiophene), 50~nm thick hole transport layers of \emph{m}-MTDATA       (4,4',4''-tris(N-3-methylphenyl-N-phenylamino)triphenylamine), and 50~nm thick electron transport layers of Bphen (4,7,-diphenyl-1,10-phenanthroline). All active layers were deposited by thermal evaporation in vacuum ($\sim$10$^{-7}$ Torr). LiF/Al cathodes (1/100~nm) deposited by vacuum evaporation through a shadow mask produce OLEDs with 3.5~mm$^2$ area. The diagram shows the relative energy level alignments and depicts the bound electron/hole complex -- the exciplex -- that forms at the interface and which gives rise to EL. In contrast to an exciton (a bound electron-hole pair residing on the \emph{same} molecule), an exciplex consists of electron and hole polarons localized predominantly on the different molecules (Bphen and MTDATA, respectively), giving the excitation a pronounced charge-transfer character.

The OLEDs were measured in vacuum at room temperature. Typical current-voltage curves exhibit the expected diode-like behavior and low turn-on voltage [inset, Fig. 1(b)].  EL spectra were measured by a 300~mm spectrometer and a charge coupled device (CCD) detector.  Fig. 1(b) shows typical EL spectra acquired in zero magnetic field, at different bias currents. The EL extends across the visible spectrum from 450-750~nm, and is peaked at $\sim$570~nm. Owing to the relative energy level alignments of MTDATA and Bphen, EL is expected to originate solely from radiative recombination of exciplexes.  The measured EL spectra are consistent with exciplex emission, and importantly, no emission is observed from excitons in the MTDATA or Bphen materials themselves (which would otherwise occur at shorter wavelengths in the 350-450 nm range \cite{Wang, YanAPL2014}), indicating the absence of unipolar currents across the heterojunction in these OLEDs. The EL spectra are in good agreement with previous studies of related exciplex-emitting OLEDs based on these materials \cite{Wang, YanAPL2014, VirgiliCPL2006, ChenAPL2013}.

Small magnetic fields \textbf{B} clearly increase the total EL emission intensity in these devices, as shown in Fig. 2(a). Figure 2(b) shows the detailed dependence of the EL intensity on \textbf{B}.  Note first the dashed black curve labeled ``full spectrum", which shows that the \emph{total} (spectrally-integrated) EL intensity grows with $|\textbf{B}|$ by $\sim$6\% at $\pm$100~mT. These overall trends are consistent with recent studies of magneto-EL phenomena in various polymer and small-molecule based OLEDs \cite{KalinowskiCPL2003, Bussman, Sheng, Iwasaki, Odaka, Wu, Desai, FJWang, BinHu, NguyenNatMat, KerstenPRL, Ehrenfreund, NguyenPRB}. The origin of this effect is believed to lie in the suppression by \textbf{B} of the mixing between singlet and triplet polaron-pairs, which form as an initial step in the exciplex formation process. Importantly, the depth and width of these magneto-EL curves shown in Fig. 2(b) provide insight into the relative rates of singlet and triplet exciplex formation, and also provide a measure of the hyperfine coupling strength between the polaron spins and the underlying nuclear spins of the molecules, as described in the following:

\begin{figure}[tbp]
\includegraphics[width=.45\textwidth]{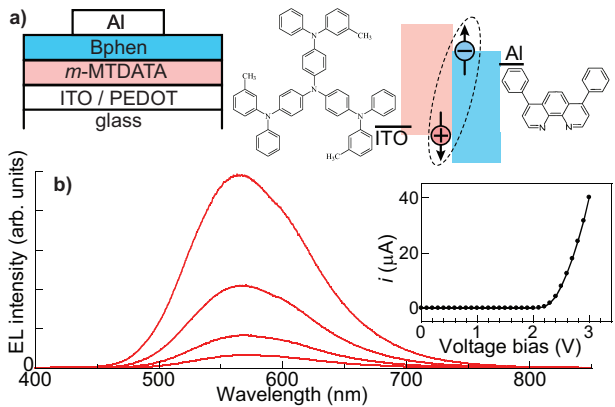}
\caption{(a) Schematic of the layered \emph{m}-MTDATA/Bphen OLEDs. The diagram depicts the relative energy level alignments and the exciplex that forms at the interface. (b) Electroluminescence (EL) spectra at 32, 16, 8, and 4 $\mu$A current bias (300~K, \textbf{B}=0). Inset: typical current-voltage behavior.} \label{fig1}
\end{figure}

These data can be understood in the context of the diagram in Fig. 2(a), which we adapt to our MTDATA/Bphen OLEDs from previous descriptions by, \emph{e.g.}, Ehrenfreund \cite{Ehrenfreund} and Kersten \cite{KerstenPRL} and the excellent discussions contained therein. In our OLEDs, free electron and hole carriers (polarons) are injected at the contacts, and these drift to the MTDATA/Bphen interface.  Coulomb attraction leads initially to the formation of very weakly bound polaron-polaron (PP) pairs, which can form as either spin-singlets or spin-triplets (PP$_\textrm{S}$, PP$_\textrm{T}$). At this stage the polarons are well-separated and electron-hole exchange interactions are negligible, so that PP$_\textrm{S}$ and PP$_\textrm{T}$ states have the same energy.

\begin{figure}[tbp]
\includegraphics[width=.45\textwidth]{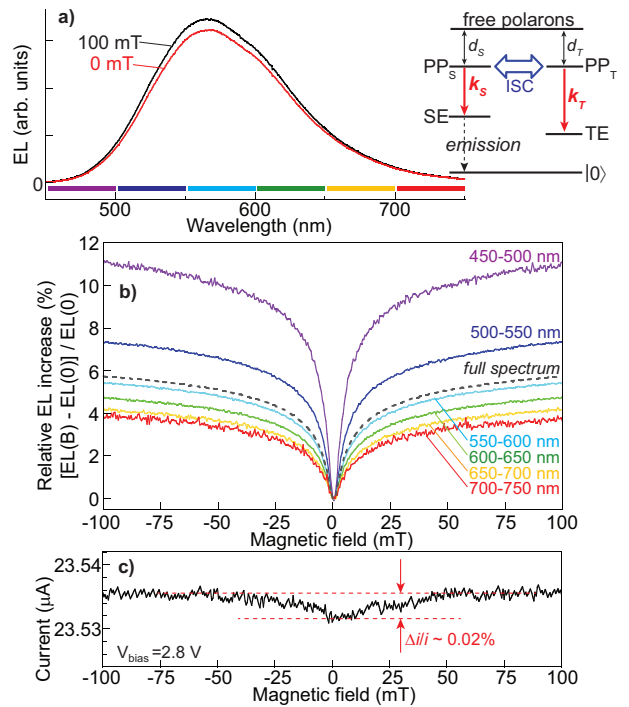}
\caption{(a) EL spectra at \textbf{B}=0 and 100~mT. Inset: Exciplex formation model. Weakly-bound singlet and triplet polaron pairs (PP) form from free carriers. PP$_\textrm{S}$-PP$_\textrm{T}$ mixing (intersystem crossing; ISC) exists at \textbf{B}=0 due to hyperfine coupling to randomly-oriented nuclear spins. Singlet and triplet exciplexes (SE, TE) subsequently form with rates $k_S$, $k_T$. Only SE recombines radiatively. (b) The relative EL intensity versus \textbf{B}, integrated and normalized over the wavelengths indicated. At blue wavelengths, the EL boost is larger (indicating larger $k_T / k_S$ ratio), and the curves are narrower (indicating weaker hyperfine coupling, due to the larger spatial extent of the polarons at these energies). The dashed line indicates the \textbf{B} dependence of the \emph{total} (spectrally-integrated) EL. (c) The OLED resistance does not vary significantly with \textbf{B}.} \label{fig2}
\end{figure}

Hyperfine coupling of the electron and hole polaron spins to the underlying `bath' of randomly-oriented nuclear spins on each molecule (primarily spin-$\frac{1}{2}$ $^1$H) leads to precession of the electron and hole spin about an effective (Overhauser) magnetic field $\textbf{B}_\textrm{hf}$. Importantly, the magnitude and orientation of $\textbf{B}_\textrm{hf}$ are different on every molecule. The independent spin precession of the electron and hole therefore causes a mixing (or `intersystem crossing', ISC) between the degenerate PP$_\textrm{S}$ and PP$_\textrm{T}$ states. From these weakly-bound states, more strongly-bound singlet and triplet exciplexes (SE and TE) can form with rates $k_S$ and $k_T$. It can be convenient to consider this last step in the exciplex-formation process as the electron (hole) polaron hopping to the Bphen (MTDATA) molecule that is \emph{closest} to the hole (electron) polaron \cite{KerstenPRL}. In this exciplex, the polarons are now sufficiently close to each other to generate a non-zero exchange splitting between SE and TE states that significantly exceeds the very small Zeeman energy from $\textbf{B}_\textrm{hf}$ ($E_Z \sim$1~$\mu$eV if $|\textbf{B}_\textrm{hf}| \sim$10~mT); therefore SE-TE mixing due to $\textbf{B}_\textrm{hf}$ does not occur.  This model assumes that only singlet exciplexes recombine radiatively.

$\textbf{B}_\textrm{hf}$ can be estimated from the sum of hyperfine energies from the randomly-oriented nuclear spins: $\sum_k a_k \mathbf{S} \cdot \mathbf{I}_k |\Psi(\mathbf{r}_k)|^2$, where $\mathbf{S}$ and $\mathbf{I}_k$ are the polaron and nuclear spin, $a_k$ is the hyperfine coupling constant, and $\Psi(\mathbf{r}_k)$ is the polaron wavefunction at the \emph{k}th nucleus. Averaging over many polarons, the root-mean-square amplitude of $\textbf{B}_\textrm{hf} \sim 1/\sqrt{N} \sim 1/\sqrt{V}$, where $N$ is the number of nuclei within the polaron wavefunction, which has characteristic volume $V$.

Crucially, if (and only if) the exciplex formation rates $k_S$ and $k_T$ are unequal, then any process that influences the PP$_\textrm{S}$-PP$_\textrm{T}$ mixing will change the balance between SE and TE populations, and will therefore affect the EL intensity. Applied fields \textbf{B} provide one such route: If $|\textbf{B}| \gg |\textbf{B}_\textrm{hf}|$, then the net magnetic fields `seen' by the electron and hole in a polaron-pair are no longer random, but are both $\approx$\textbf{B}. Electron and hole polarons therefore precess in synchrony (provided their $g$-factors are similar), and PP$_\textrm{S}$-PP$_\textrm{T}$ mixing is suppressed.  The characteristic half-widths of the magneto-EL curves in Fig. 2(b) therefore indicate when $|\textbf{B}|\sim |\textbf{B}_\textrm{hf}|$, providing a measure of the strength of hyperfine coupling in these molecules.  Moreover, the \emph{depth} of the curves provides insight into the ratio of exciplex formation rates $k_T/k_S$.

Magneto-EL studies to date have considered only the field dependence of the \emph{total} (\emph{i.e.}, spectrally-integrated) EL intensity \cite{KalinowskiCPL2003, Bussman, Sheng, Iwasaki, Odaka, Wu, Desai, FJWang, BinHu, NguyenNatMat, KerstenPRL, Ehrenfreund, NguyenPRB}, which is shown here for our OLEDs by the dashed line in Fig. 2(b).  However, our spectrally-resolved measurements reveal a surprising and significant difference in both the depth and width of the magneto-EL data as a function of emission wavelength.  As shown in Fig. 2(b), \textbf{B} induces a much larger EL boost at the blue end of the EL spectrum ($\sim$11\%) as compared to the red end ($\sim$4\%). (These values remain unchanged at higher \textbf{B} to 300~mT, and are independent of field direction.)  It is also clearly evident that the \emph{widths} of the magneto-EL curves vary significantly across the EL spectrum. The curves are narrower at blue wavelengths, indicating smaller hyperfine fields $\textbf{B}_\textrm{hf}$ for exciplexes that form at high energies. Since the rms magnitude of \textbf{B}$_\textrm{hf}$ $\propto 1/\sqrt{V}$ as discussed above, this directly provides information about the localization and effective `size' of the underlying polaron wavefunctions.

We note that the large EL changes are not simply due to magneto-resistive effects that can occur in organic materials \cite{Wagemans, Harmon}.  Figure 2(c) shows the simultaneously-measured current through the OLED, which changes only slightly (0.02\%). Therefore, to leading order \textbf{B} influences primarily the radiative exciplex recombination rate, but not the total recombination rate.

\begin{figure}[tbp]
\includegraphics[width=.44\textwidth]{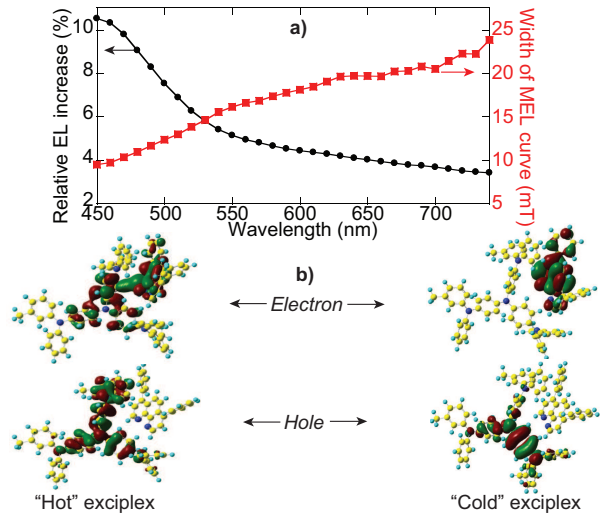}
\caption{(a) Left axis (circles): The relative increase in EL intensity due to \textbf{B} (\emph{i.e.}, the `depth' of the curves shown in Fig. 2b), versus EL wavelength. Short (blue) wavelengths show the largest change indicating larger $k_T/k_S$ ratios. Right axis (squares): The corresponding full width of the magneto-EL curves ($\propto 2|\textbf{B}_\textrm{hf}|$). $\textbf{B}_\textrm{hf}$ is smaller at blue wavelengths, indicating more delocalized (larger) polarons. (b) Natural Transition Orbitals of an electron and hole, calculated for `hot' (unrelaxed) and `cold' (vibrationally relaxed) exciplexes on a MTDATA/Bphen molecular pair.} \label{fig3}
\end{figure}

Figure 3(a) summarizes these trends by showing both the spectrally-resolved depth and width of the magneto-EL data. The magnitude of the effect (the depth) drops rapidly across the blue-green part of the EL spectrum (from 450 to 550 nm), and eventually falls to one-third of its maximum value by 750~nm. This trend indicates that the exciplex formation ratio $k_T/k_S$ decreases towards unity across the spectrum, likely due in part to a shrinking SE-TE exchange splitting as discussed below. Concurrently, the width of the magneto-EL curves increases steadily across the spectrum and approximately doubles in value, suggesting that the characteristic volume $V$ of the polaron wavefunctions in low-energy exciplexes has shrunk four-fold.

To visualize the polaron wavefunctions, we performed electronic structure calculations of a Bphen/MTDATA molecular pair. An optimal ground state geometry was obtained using Density Functional Theory (DFT), and electronic excitations and excited geometries were calculated using time-dependent DFT \cite{DFT, Chai, Grimme, Magyar}. Figure 3 shows natural transition orbitals \cite{Martin} of an electron and hole for an unrelaxed (higher energy, or ``hot") exciplex and also for a vibrationally relaxed (lower energy, or ``cold") exciplex. Importantly, in the former case the electron and hole wavefunctions are more spatially extended and have some overlap due to partial delocalization of the electron between Bphen and MTDATA, which will generate larger SE-TE splitting (estimates of the exchange energy range from 10-100 meV given the uncertainty of molecular orientations and the choice of DFT model). In the latter case the wavefunctions are more localized on their respective molecular species (with minimal overlap, so that SE and TE states are closer in energy and $k_T/k_S \rightarrow$1).

\begin{figure}[tbp]
\includegraphics[width=.40\textwidth]{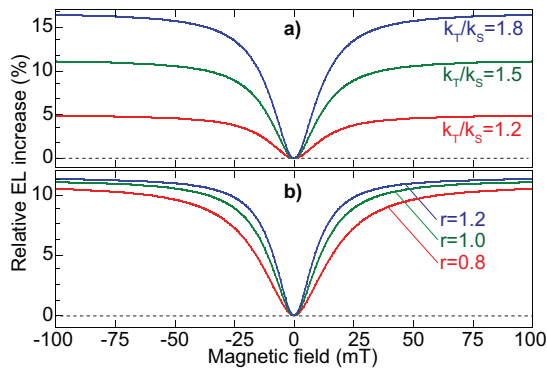}
\caption{Modeling the magneto-EL. The key parameters are the ratio of exciplex formation rates $k_T/k_S$, and the normalized lengthscale $r$ of the polaron excitation. The characteristic hyperfine field $\textbf{B}_\textrm{hf} \propto 1/\sqrt{V} \propto r^{-3/2}$.} \label{fig4}
\end{figure}

To model the magneto-EL data analytically, we employed a density matrix model for the ensemble of electron and hole polaron spins in the PP states, including Zeeman terms and hyperfine interactions with randomly-oriented $^1$H nuclei. The time evolution of the density matrix is described by a stochastic Liouville equation \cite{KerstenPRL}.  Steady-state analytical solutions are obtained using Bloch-Wangsness-Redfield theory \cite{MRBook}. As polarons hop between different molecules they experience fluctuating $\textbf{B}_\textrm{hf}$. The resulting SE and TE densities are calculated from the corresponding PP density matrix elements and the formation rates $k_T$ and $k_S$.

Two key parameters control the shape of the calculated magneto-EL curves, which are determined from the relative SE density: the ratio $k=k_T/k_S$, and the spatial extent of the polarons as characterized by a normalized length $r$.  Both parameters vary locally due to random steric interactions between the molecules. The model allows for different forms of the spin-spin correlation function with the simplest being a single exponential decay, which yields magneto-EL curves with inverted Lorentzian lineshape when $k>1$. In Fig. 4(a) $k$ is varied while $r$ is fixed, and in Fig. 4(b) $r$ is varied while $k$ is fixed.  The `depth' of the magneto-EL curves is primarily determined by $k$, and the width by $r$.  More localized polarons (smaller $r$, and therefore larger $\textbf{B}_\textrm{hf}$) are associated with EL at longer wavelengths, while delocalized polarons are associated with EL at shorter wavelength. As seen in Fig. 3(b), the former have smaller electron-hole overlap and therefore smaller SE-TE splitting. Thus $k$ is smaller (closer to unity) than those corresponding to the less tightly bound polarons.  The calculated line shapes and trends are very consistent with experimental data.

In summary, spectrally-resolved magneto-EL measurements provide considerable insight into the energy-dependent exciplex formation rates and hyperfine interactions in OLEDs, both of which depend in turn on the localization and detailed spatial extent of the electron and hole polarons themselves. We anticipate that similarly valuable information can be obtained for situations where light emission derives from excitons in single-component OLEDs. This work was supported by the Los Alamos LDRD program.


\begin{references}
\bibitem{DFT}We used wB97XD/6-31G* model chemistry. wB97XD is a range-corrected functional with empirical London dispersion terms, allowing to describe excitons, charge-transfer states, and weakly van-der-Waals bound complexes. Simulations used Gaussian 09 software.
\end{references}
\end{document}